\documentclass{article}
\usepackage{arxiv}
\usepackage[utf8]{inputenc} % allow utf-8 input
\usepackage[T1]{fontenc}    % use 8-bit T1 fonts
\usepackage{hyperref}       % hyperlinks
\usepackage{url}            % simple URL typesetting
\usepackage{booktabs}       % professional-quality tables
\usepackage{amsfonts}       % blackboard math symbols
\usepackage{nicefrac}       % compact symbols for 1/2, etc.
\usepackage{microtype}      % microtypography
\usepackage{lipsum}
\usepackage{graphicx}
\graphicspath{ {./images/} }
\usepackage{float}
\usepackage{CJKutf8}
\newcommand{\commentword}[1]{}

\newpage
\title{Amplifying vortex shedding for energy harvesting with active flow control}
\author{
Varun Varma Jaganath \\ 
  School of Mechanical Sciences\\
  Indian Institute of Technology\\
  Bhubaneswar, India 752050 \\
  \texttt{21me01018@iitbbs.ac.in} \\
     \And
Ben Steinfurth \\ 
  Institute of Aeronautics and Astronautics \\
  Technische Universität Berlin\\
  10623 Berlin, Germany \\
  \texttt{ben.steinfurth@tu-berlin.de} \\
}

\begin{document}
\maketitle
\begin{abstract}
Energy harvesting from vortex-induced vibrations is a promising technology that relies on the vibrations of bluff bodies due to vortex shedding. Increasing the vibration amplitude at a given free stream kinetic energy is therefore equivalent to enhancing the efficiency of the harvesting device. In this study, we assess the potential of alternate slot blowing to amplify force fluctuations. Pressurized air is ejected alternatingly from the top and bottom parts of the cylinder. Through experimentation in a low-speed wind tunnel ($Re=8,000$), we show that the magnitude of lift fluctuations can be enhanced by up to a factor of three compared to the unforced flow when the actuation is aligned with the natural vortex shedding frequency. Velocity field measurements indicate that this is caused by strong streamline bending whereas, at a higher forcing frequency, vortex shedding is suppressed. The results presented in this article suggest that a significant increase in the dynamic load acting on a cylinder can be achieved with carefully chosen active flow control parameters, thereby promoting future energy harvesting applications.
\end{abstract}


\section{Introduction}
The flow of fluids around structures, ranging from the Aeolian Harp \cite{Yu2022} to skyscrapers \cite{Yan2022}, can cause fluctuating aerodynamic loads. Historically, the resulting vibrations have always been interpreted as destructive, for instance in the construction of ocean pipelines \cite{chaudhury2001vortex}, oil-refinery support structures, heat exchanger tubes \cite{islam2023flow}, and cable lines \cite{Blevins2001}. Since the collapse of the Tacoma Narrows Bridge in 1940, the idea of flow-induced vibrations has been well studied and documented under the umbrella term of bluff body flows or fluid-structure interaction problems \cite{WILLIAMSON2008713}.

Beyond the critical Reynolds number $Re = 47$ \cite{lienhard1966synopsis}, in the case of circular cylinders, the Kelvin-Helmholtz shear instabilities grow which is manifested in the ejection of two symmetric vortices of opposite vorticity. These vortices are shed periodically at a characteristic Strouhal number \cite{Strouhal1878} depending on the geometry of the bluff body. What is described here is the famous vortex street \commentword{Theodore von Karman (1911)} \cite{karman1912mechanismus}, which is understood to represent the stable limit cycle behaviour of the dynamical system \cite{Landau1970}.

As mentioned, the vibrations due to the vortex street may cause large-scale destructive interactions. Suppression of the vortex street has been extensively researched by active and passive means aiming to reduce vortex-induced vibrations (VIVs). Passive flow control involves no external energy addition to the flow but uses structural modifications or auxiliary wake suppression devices. Based on analytical formulations given in Ref. \cite{blevins1976fluid} or experimental studies presented in Ref. \commentword{Feng (1968)} \cite{feng1968measurement}, increased damping is an effective approach, up to the point that vibration amplitudes can be assumed negligible when the reduced damping factor exceeds 64. Other passive techniques involve streamlining the cross-section \commentword{Tobes et al (1968)} \cite{Toebes1969}. This approach is most effective when the flow direction is fixed and the structure has sufficient stiffness to overcome flutter. Wake suppression can also take place through helical strakes \commentword{Wilson and Tinsley (1989)} \cite{wilson1989vortex}, axial slats \commentword{Wong et al. (1982)} \cite{wong1982comparative} and splitter plates \commentword{Sallet (1970)} \cite{sallet1970method}, all of which perturb the shear layer or counter the adverse pressure gradient and thereby effectively suppress vibrations. Passive enhancements are, however, non-dynamic and hence could remain ineffective if the Reynolds number of the stream changes.

Active flow control (AFC), as a means of preventing separation, has traditionally been associated with the injection of fluid to, or the removal of fluid from, a boundary layer \cite{flatt1961history, Schlichting1968}. The term active refers to an input of external energy to control flow characteristics such as separation delay and vortex shedding suppression. Excitation is often realized as an oscillatory addition of momentum \cite{seifert1998use}. Historically, however, steady suction and blowing were the first to be implemented and offered the advantage of controllability and maneuverability \cite{article}. The basic principle of suction is to remove low-momentum fluid near a surface and deflect the high-momentum free-stream fluid toward the surface. Blowing involves the idea of providing additional momentum to the flow, energizing the boundary layer, and giving the flow additional resistance to separation. As for the $Re > 47$ cylinder flow, a slight blowing or sufficiently high suction pressure at the base stablizes the wake and reduces the instability \commentword{Delaunay et al, 2001} \cite{delaunay2001control}.  In Ref. \commentword{Bhattacharya and Gregory (2018)} \cite{bhattacharya2018optimum}, it is observed that near-wake longitudinal vortices at an optimal wavelength can suppress vortex shedding and nullify vibrations. Unsteady blowing is often facilitated using synthetic jets \commentword{Glezer and Amitay (2002)}\cite{glezer2002synthetic}. Dielectric barrier discharge (DBD) plasma actuators have also gained popularity because they do not have moving parts. 

In the recent past, the necessity to harvest energy from renewable resources has emerged into importance with the existing sources of fossil fuel-driven power coming to an end. Wind and ocean flows, seemingly ubiquitous, give immense prospectus for a plentiful source of clean and renewable energy through waves, currents, tides, and airflows.

At the microscale, the continuous development of low-power consumption sensors, wireless networking technology, and micro–nano manufacturing technology has led to the extensive application of wireless sensor networks in various industries. With the advent of the Internet of Things (IoT), the problem of power supply for wireless sensor nodes in the network has attracted wide attention. Traditionally, there are some serious limitations of the battery power supply, such as the large size, limited life, regular replacement or charging, and environmental pollution \cite{lain2021understanding}. Hence, small-scale energy harvesting techniques have the potential to substitute batteries. To this end, piezoelectric energy harvesting from VIVs has been a recent development to facilitate small-scale needs such as powering wireless sensor nodes on an IoT network or recharging batteries in remote areas and can be further extended to power micro-electro-mechanical systems \cite{lu2022wind}. Various structures have been explored such as piezoelectric leaves in the wake of the cylinder \commentword{Li et al (2009)} \cite{li2009vertical}, bluff bodies  located at the edge of piezoelectric cantilevers \commentword{Sun and Seok (2020)} \cite{sun2020novel}, and cylinders located in the wake of the cantilever disjoint from each other \commentword{Zhang et al (2017)} \cite{zhang2017improving}, enhancing the efficiency and bandwidth of the harvester. Further, non-linearly coupled magnetic systems with internal resonance features have also been explored \commentword{Liuyang et al (2016)} \cite{xiong2016internal}.  

The energy harvesting systems introduced above, though promising, require the presence of a strong vortex-shedding interaction to ensure large amplitude VIVs. This condition is not met at all times, specifically when the vortex shedding does not excite the system resonance.
To overcome this shortcoming, AFC has been proposed for the benefit of VIV-based energy harvesting, for instance in Ref. \commentword{Greenblatt (2018)} \cite{garzozi2018pulsed}. Investing external energy implies that the the amount of energy retrieved should be beyond a practical threshold. In this context, the efficacy of pulsed blowing has been explored by analyzing the non-linearities of such AFC systems, yielding the required momentum to be added to flow and a model of the maximum theorized power that can be harnessed through this system. Computational work in Ref. \commentword{Gao et. Al (2023)} \cite{guo2023effects} presented the various vortex interactions that take place during pulsed blowing. As an alternative to the ejection of compressed air, DBD actuators may also be used to enhance lift fluctuations \cite{garzozi2023wind}, \cite{greenblatt2012flow}.

Considering the recent progress of VIV-based energy-harvesting devices, but also their limitations in terms of power output, we aim to shed some light on the flow physics that govern the amplification of vortex shedding through AFC. Specifically, we attempt to identify effective forcing time scales and enable novel insights into the time-dependent velocity field of a controlled cylinder flow.

The remainder of this article is structured as follows: after explaining the experimental approach in the next section, wall pressure and velocity field measurements will be presented in Sec. 3. Then, in Sec. 4, some final remarks will be offered.

\section{Methods}
\subsection{Experiment description}
\label{sec:headings}
Experiments were performed in a low-speed blowdown wind tunnel with a test section of $400\,\mathrm{mm}$ both in height and in span.  The side walls of the test section are made of acrylic glass to enable optical access. The $D=80\,\mathrm{mm}$ cylinder was mounted horizontally between the side walls. The maximum wind speed that can be achieved in the test section is $45\,\mathrm{m/s}$ with a free stream turbulence intensity of less than 1\%. The wind tunnel was operated at two inflow velocities $U_\infty = (1.6, 10)\,\mathrm{m/s}$, corresponding to Reynolds numbers $Re = U_\infty D/\nu \approx (8,000 , 50,000)$, which is the flow regime for laminar boundary layer development on the cylinder \cite{lienhard1966synopsis}.\\

Alternate slot blowing (ASB) was chosen as the technique to manipulate the temporal evolution of the aerodynamic forces. Following this approach introduced in Ref. \cite{garzozi2023wind}, pulsed jets are ejected alternatingly from slots in the top and bottom part of the cylinder, respectively. This way, flow separation is suppressed periodically for short amounts of time.
 
\begin{figure}[H]
  \centering
  \includegraphics[width=.7\textwidth]{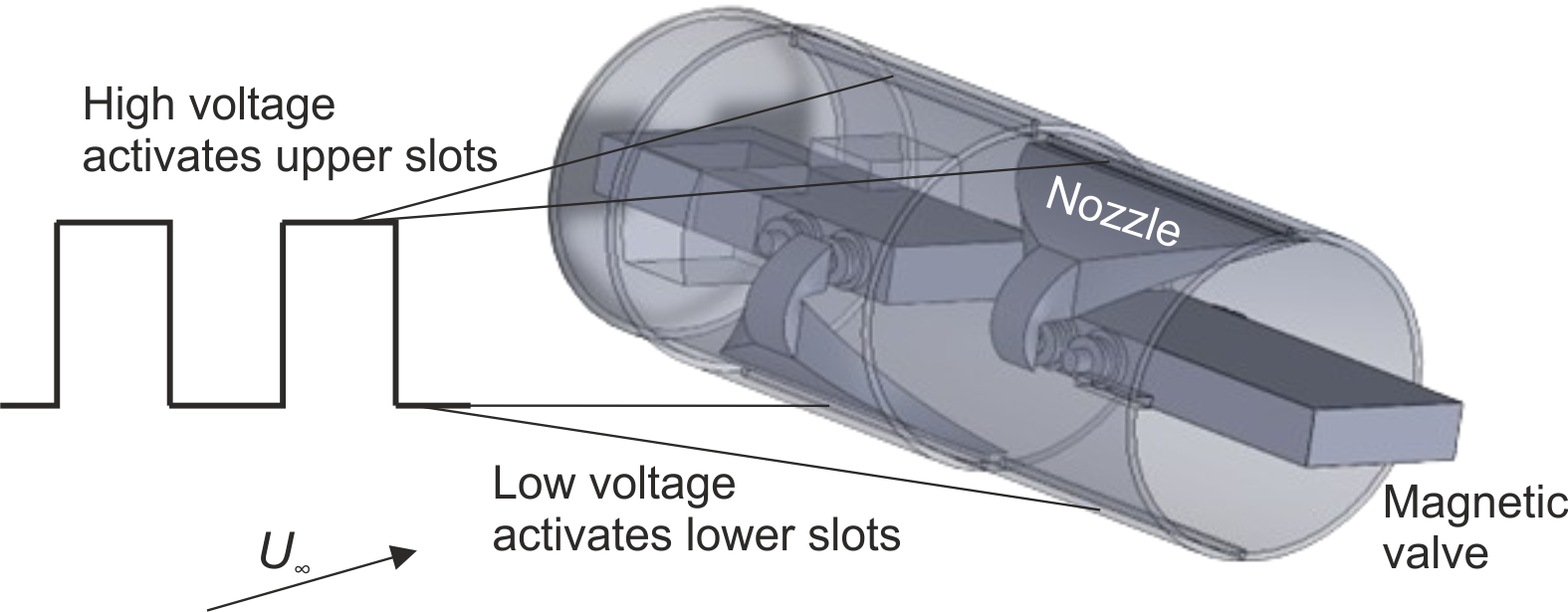}
 \caption{Isometric view of hollow cylinder with magnetic valves integrated (only one nozzle shown for each of the two spanwise locations)}
  \label{fig:image2}
\end{figure}

 In the present study, four nozzles (at two spanwise stations) with outlet dimensions of $80 \times 1 \, \mathrm{mm^2}$ were designed to feed compressed air into the boundary layer (Fig.~\ref{fig:image2}). The pulsation was realized using magnetic valves with a maximum switching frequency of $f=300\,\mathrm{Hz}$, covering a certain spanwise extent on both sides of the cylinder symmetry plane. Each valve had two output ports connected to slots located at 90 and 270 degrees respectively. Depending on the current voltage supplied to the valve controller, which followed a square-wave signal (Fig.~\ref{fig:image2}), compressed air was either ejected through the top or the bottom slot. The jet emission angle was $\varphi = 30^\circ$ with respect to the inflow direction, representing an effective operation condition for separation control \cite{Steinfurth2021}.

%\begin{figure}[H]
%  \centering
% \includegraphics{image3.png}
%  \caption{Mathematical Representation of ASB}
%  \label{fig:image3}
%\end{figure}

Jet velocities were regulated by a mass flow controller with a range of up to $\dot{m}\approx 8\cdot 10^{-4}\,\mathrm{kg/s}$ and a low uncertainty of $\pm 0.1\,\%$ full-scale and $\pm 0.5\,\%$ of the measured value. The forcing intensity is quantified by the velocity ratio between jet and free stream, and the ratios $V_\mathrm{R} = u_\mathrm{jet} / U_\infty \approx (0.4, 0.6, 1.0, 1.3, 1.9, 2.2)$ were investigated, corresponding to momentum coefficients in the range $c_\mu = 2 \cdot \frac{h u_\mathrm{jet}^2}{D U_\infty ^2} = 0.4\%, \dots, 11\%$ ($h=1\,\mathrm{mm}$ is the slot width).\\

\begin{figure}[h]
  \centering
  \includegraphics{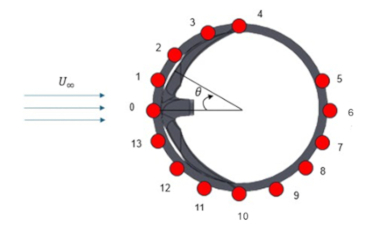}
  \caption{Cross-section view inside symmetry plane, red circles indicate positions of pressure taps}
  \label{fig:image1}
\end{figure}

To assess the effect of forcing on the fluctuation of aerodynamic forces, the cylinder featured a number of pressure taps inside the symmetry plane (i.e., in-between the jet outlets, see Fig.~\ref{fig:image1}).

 After validating that the pressure distribution was symmetrical, only the pressure taps located on the lower half of the cylinder were evaluated to compute integral forces. Differential piezoelectric pressure transducers with a range of $p=\pm 250\,\mathrm{Pa}$ and a sensitivity of $S=\pm 8\,\mathrm{mV/Pa}$ were integrated into the hollow cylinder. These sensors measured the differential pressure between the local wall pressure and the free stream static pressure to obtain the difference in static pressure. The free stream static pressure $p_\infty$ was obtained from a pitot-static tube installed at the upstream end of the test section. The sensors were sampled at an acquisition rate of 1000 Hz, and each set of data was obtained by running the experiments for at least 100 cycles of the characteristic dynamical event (i.e., periodic forcing or vortex shedding).

In addition to pressure measurements, phase-locked PIV was performed to determine the time-resolved two-component velocity field inside the cylinder symmetry plane. An Nd:YAG dual-cavity laser was used to illuminate aerosol particles supplied to the flow, and the tracer reflections were recorded with a CMOS $5\,\mathrm{MP}$ camera.  The square-wave signal driving the magnetic valves was used to trigger the PIV system at 20 pre-defined phases spanning the ASB forcing period; 100 snapshots were recorded for each phase ensuring convergence of the phase-averaged velocity fields. Image pre-processing involved minimum-intensity background subtraction before cross-correlation with iterative grid refinement was applied.

\subsection{Baseline cylinder flow (without AFC)}
To validate the setup, the mean pressure distribution and the vortex shedding characteristics are compared against available data from the literature in the following.

\begin{figure}[H]
  \centering
  \includegraphics{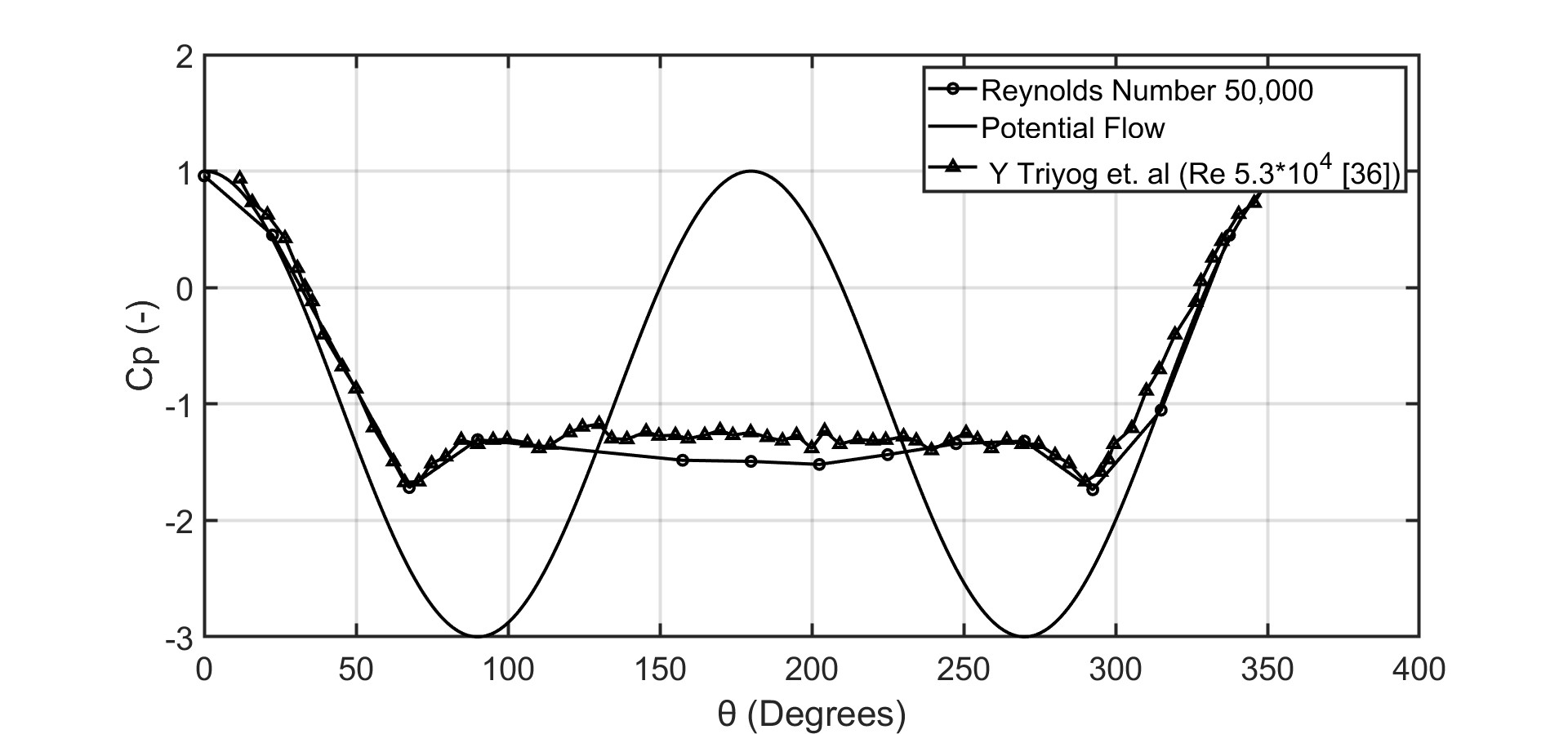}
  \caption{Mean pressure distributions at $Re=50,000$ in the absence of AFC, compared with experimental data obtained from Ref. \cite{triyogi2009reducing}}
  \label{fig:image5}
\end{figure}

The time-averaged pressure distribution obtained with our experiment is presented in Fig.~\ref{fig:image5} (filled circles) and compared with data from Ref. \cite{triyogi2009reducing}. Aside from some deviation with regards to the base pressure ($\theta = 120^\circ, ..., 240^\circ$), the distributions are in good agreement. Hence, the present set-up is deemed well suited to determine the mean aerodynamic loads.

%\begin{figure}[H]
  %\centering
  %\includegraphics{image6.png}
  %\caption{}
 % \end{figure}

\begin{figure}[H]
\centering
\includegraphics{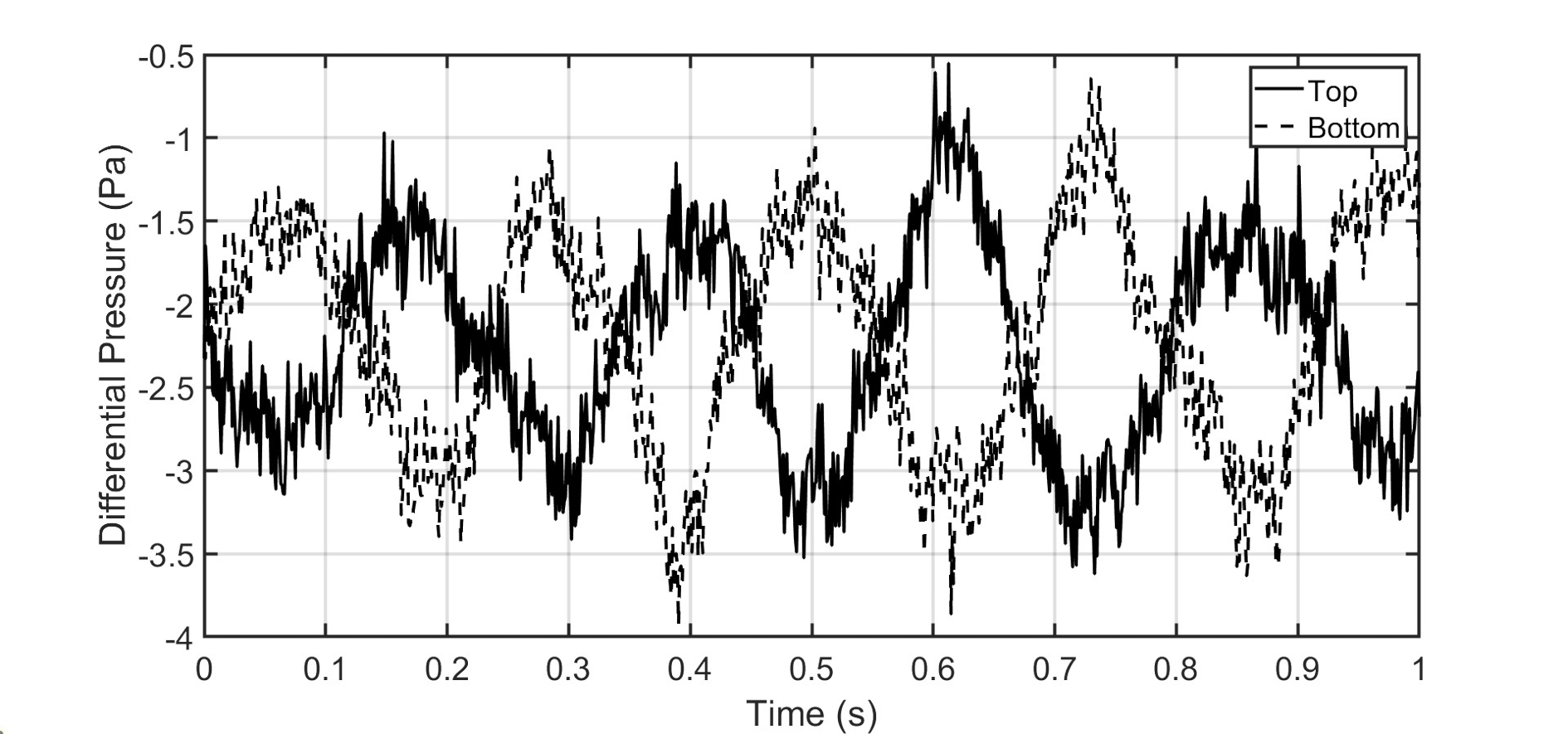}
\caption{Time series of wall pressure on opposite cylinder sides for $Re = 8,000$}
\label{fig:image5b}
\end{figure}

The time-varying static pressure difference for two sensors located at the top and bottom are presented in Fig.~\ref{fig:image5b}. As expected, an anti-correlation is found: high values at the top of the cylinder are accompanied by low values at the bottom of approximately the same magnitude, and vice versa. This justifies the usage of only pressure taps on the bottom side to evaluate lift fluctuations. The frequency associated with the displayed fluctuations corresponds to $St \approx 0.2$.

From the measured time-dependent pressure distributions, lift coefficients are determined using trapezoidal integration. The root-mean-square (rms) value associated with lift fluctuations is $c'_\mathrm{l, rms}\approx 0.22$ which falls slightly below the value of $c'_\mathrm{l, rms}\approx 0.26$ reported in \cite{garzozi2023wind}.

%\begin{figure}[H]
%\centering
%\includegraphics{image5a.png} 
%\caption*{Figure 5a: Time series of wall pressure on opposite cylinder sides for Re 50,000 }
%\label{fig:image5a}
%\end{figure}

In summary, both the mean pressure distribution and the force fluctuations observed in our experiment are in agreement with previous literature results. It is therefore reasonable to believe that findings regarding the application of AFC, presented in the next section, are generalizable.

\section{Results}

The main focus in the following section lies on the effect of AFC on vortex shedding. Specifically, three different reduced forcing frequencies $F^+=(0.03, 1.12, 0.63)$ are assessed, where $F^+=fD/U_\infty$ is the reduced forcing frequency based on the dimensional pulsation frequency $f$. We restrict ourselves to the Reynolds number $Re = 8,000$ in the following.

\subsection{Mean load analysis}
Figure 6 shows the pressure coefficients at a velocity ratio $V_\mathrm{R}=2.2$ with varying actuation frequencies.

\begin{figure}[H]
 \centering
 \includegraphics{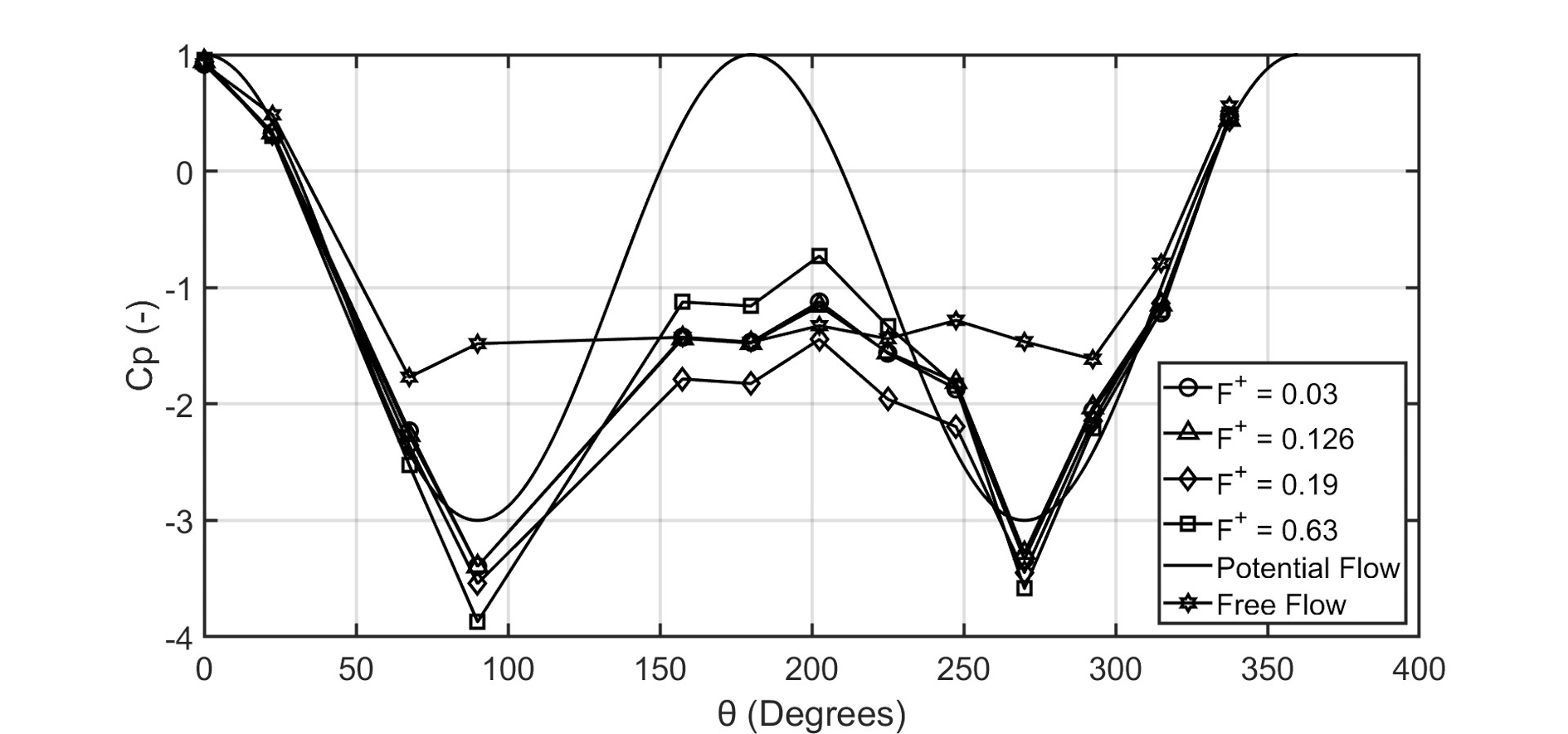}
 \caption{Mean pressure distributions for different forcing frequencies at $V_\mathrm{R}$= 2.2 and  $Re=8,000$}
 \label{fig:image10}
\end{figure}

Here, the dependency of the AFC effectiveness on the frequency of pulsation is evident. Across all the frequencies, a localized low pressure is created at the exit of the slots at 90 and 270 degrees. This is partly due to the stronger streamline bending (separation control), but considering that the pressure drop is stronger than in the case of the inviscid solution, AFC also appears to yield circulation enhancement.

Interestingly, the magnitude of base pressure depends on the frequency of pulsing. Specifically, the two lowest frequencies of $F^+ = 0.03$ (circles)  and $F^+= 0.13$ (triangles) do not yield significant effects. In contrast, the base pressure is reduced at $F^+=0.19$ and increased at $F^+=0.63$. 

We conclude that the mean aerodynamic load, especially drag, is sensitive towards the choice of the forcing frequency.\\

\subsection{Dynamically forced flow field}

\begin{figure}[H]
    \centering
    \includegraphics[width=0.8\textwidth]{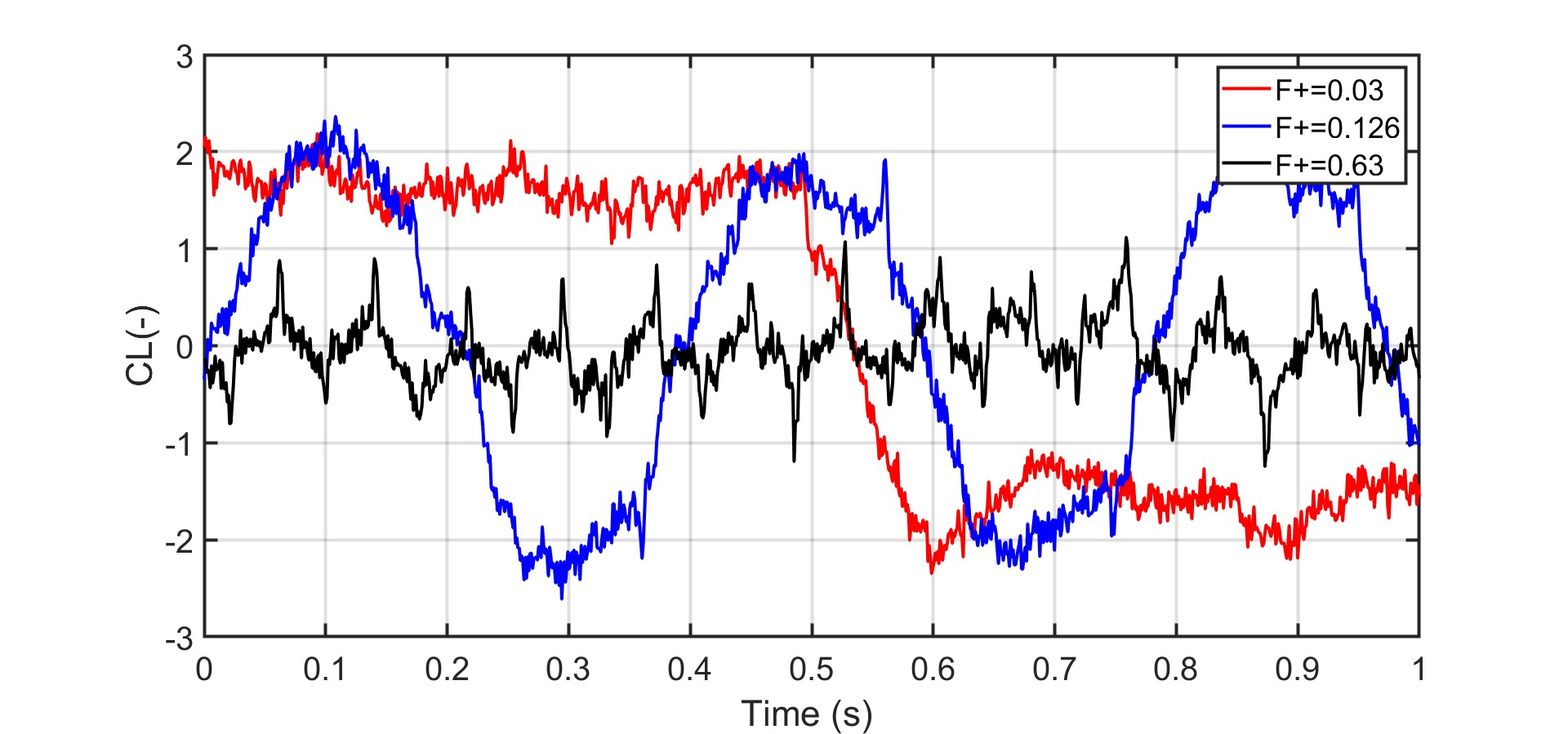}
    \caption{Lift coefficient time series for different forcing frequencies}
    \label{fig:image13}
\end{figure}

In Fig. \ref{fig:image13}, the temporal evolution of the lift coefficient is presented for three selected forcing frequencies, corresponding to the sub- and super-resonance regime ($F^+=0.03$ and $F^+=0.63$) whereas the third frequency $F^+=0.13$ is close to the natural vortex shedding frequency.

It is immediately apparent that large lift fluctuations are only caused by the two smaller forcing frequencies, yielding lift coefficients that oscillate in the range $c_l = -2, \dots , 2$. The fluctuations observed for $F^+=0.63$ are only half as large.

\begin{figure}[H]
 \centering
 \includegraphics[width=1\textwidth]{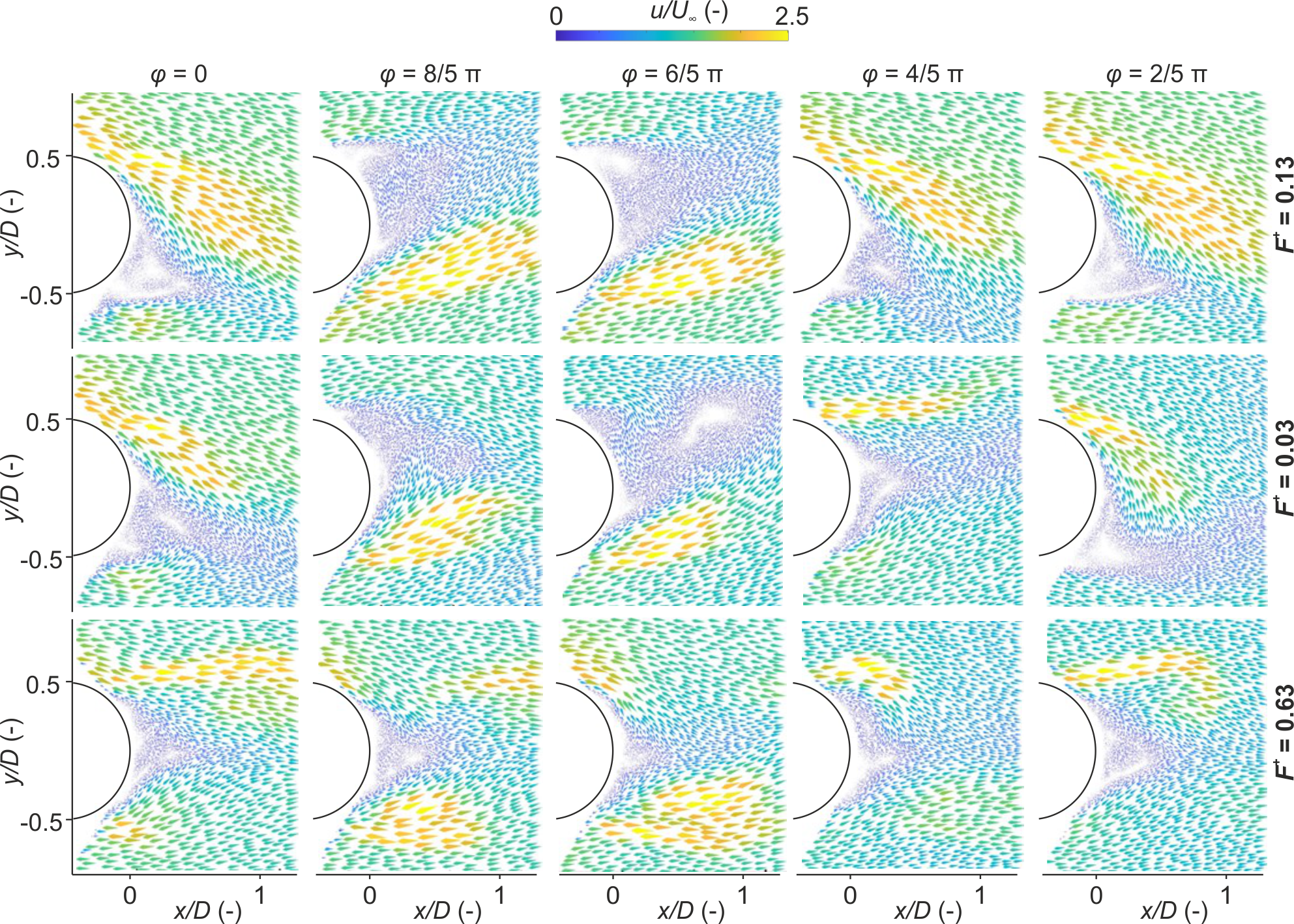}
 \caption{Phase-averaged velocity fields for different ASB frequencies (row-wise)}
 \label{fig:image12}
\end{figure}

To relate the information regarding lift fluctuations to the flow field, selected phases during the ASB cycle are presented in Fig.~\ref{fig:image12} which represents, to the best of our knowledge, the first flow field analysis of ASB. Across all cases, the ochre-yellow hue indicates high velocities induced by the jet emission and the blue color highlights the reverse-flow inside the cylinder wake.

The top row shows velocity fields for $F^+=0.03$ (sub-resonance forcing). At the first displayed phase, the upper actuation slot is active, delaying separation on the upper part of the cylinder and deflecting the wake towards the other side. For the next two phases ($\varphi = 2/5 \pi$ and $\varphi = 4/5 \pi$), it is the other way around. Due to the relatively long forcing period, the region of high velocity covers a substantial streamwise extent.

At $F^+=0.13$, the forcing frequency is closer to the characteristic vortex shedding frequency, hence stronger interactions between ASB and the natural flow can be expected.

Finally, at $F^+=0.63$ (super-resonance regime), the high-frequency pulsation clearly reduces the region of reverse-flow to the greatest extent and therefore caters best to pressure drag reduction. However, the streamline bending is not as strong as for the other forcing frequencies as the wake region retains an almost symmetric shape throughout the actuation period.

\subsection{Amplification of dynamic loads}
\begin{figure}[H]
\centering
\includegraphics[width=0.8\textwidth]{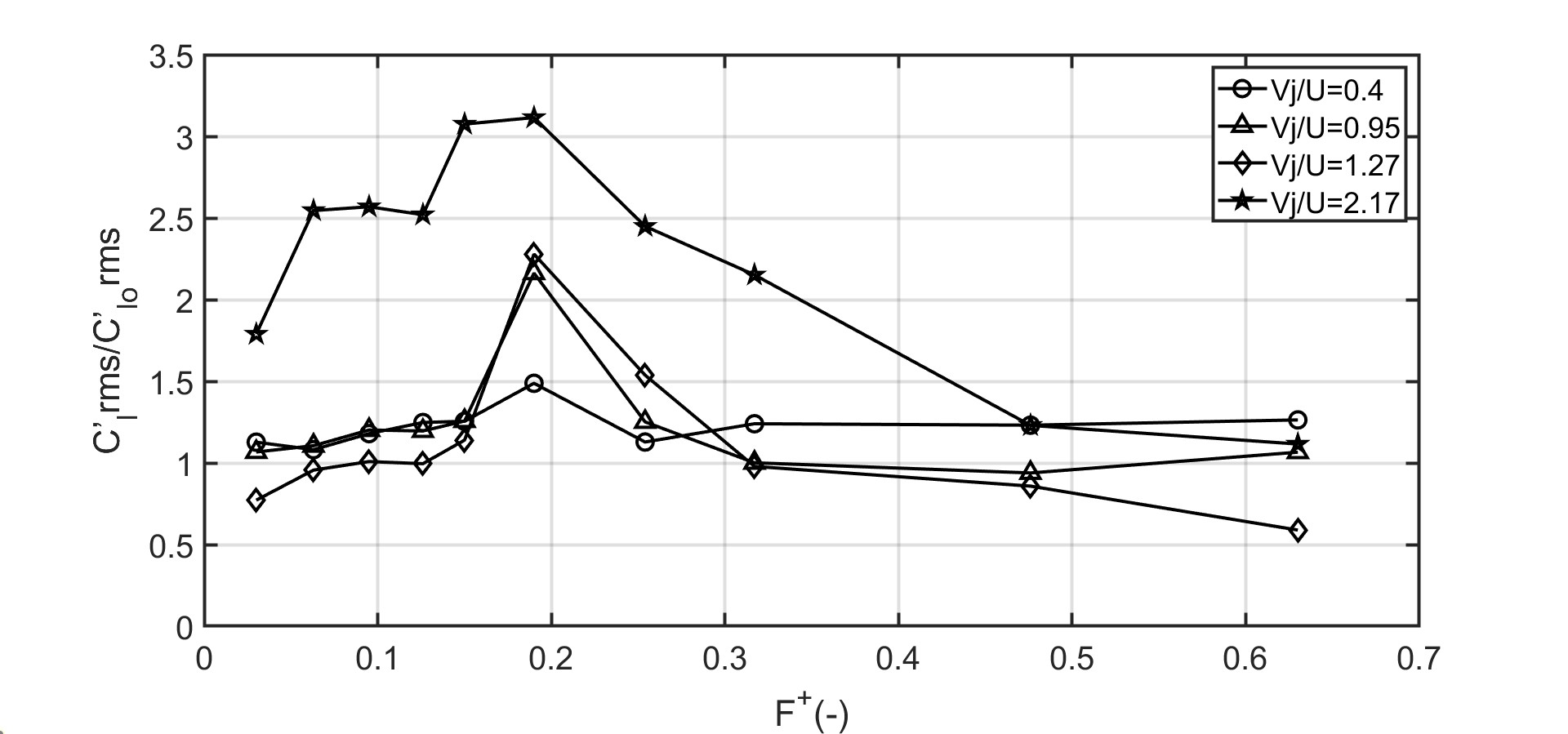}
\caption{Gain in lift fluctuations depending on forcing frequency and velocity ratio}
\label{fig:image15}
\end{figure}

After assessing some selected AFC cases in detail, we now present an overview regarding lift fluctuations achieved with ASB. Recall that their significance lies in the fact that they determine the amplitude of vibrations and thereby the energy output of future energy harvesting devices. The RMS values of lift fluctuations are presented in Fig.~\ref{fig:image15}. Note that they are normalized with the value found for the natural flow field (without ASB).

As indicated above, the lift fluctuations are maximum for the forcing frequencies close to the natural vortex shedding frequency ($St = 0.2$). Here, all velocity ratios yield beneficial effects, ranging from increases by a factor of $1.5$ (low-amplitude forcing, $V_\mathrm{R}=0.4$) to a factor of three at $V_\mathrm{R}=2.2$. As expected, the magnitude of lift fluctuations is observed to be proportional to the amount of momentum added to the flow. However, as one moves away from the vortex shedding frequency, the enhancement of lift fluctuations becomes negligible.

\subsection{Frequency lock-in effect}
By analyzing the spectral response of the various lift signals for different parameter combinations, the dominant mode between ASB and natural vortex shedding can be identified. The general trend is as follows (Fig 10): at the lowest velocity ratios ($V_\mathrm{R}<0.4$), the dominant dynamical event is natural vortex shedding since, independent of the forcing frequency, the highest fluctuation amplitude is found at $St \approx 0.2$. As we move to higher velocity ratios ($V_\mathrm{R}=1.3$), the ASB pulsing begins to dominate the flow - but, importantly, only in a spectral range close to the natural vortex shedding. For the largest investigated velocity ratio ($V_\mathrm{R}=2.2$), the dominant frequency always corresponds to forcing. In other words, the dynamics of the flow field are governed by ASB. Yet, it is important to retain in mind that the largest lift fluctuations are observed when the frequency is close to the natural shedding frequency.

\begin{figure}[H]
    \centering
    \includegraphics[width=0.8\textwidth]{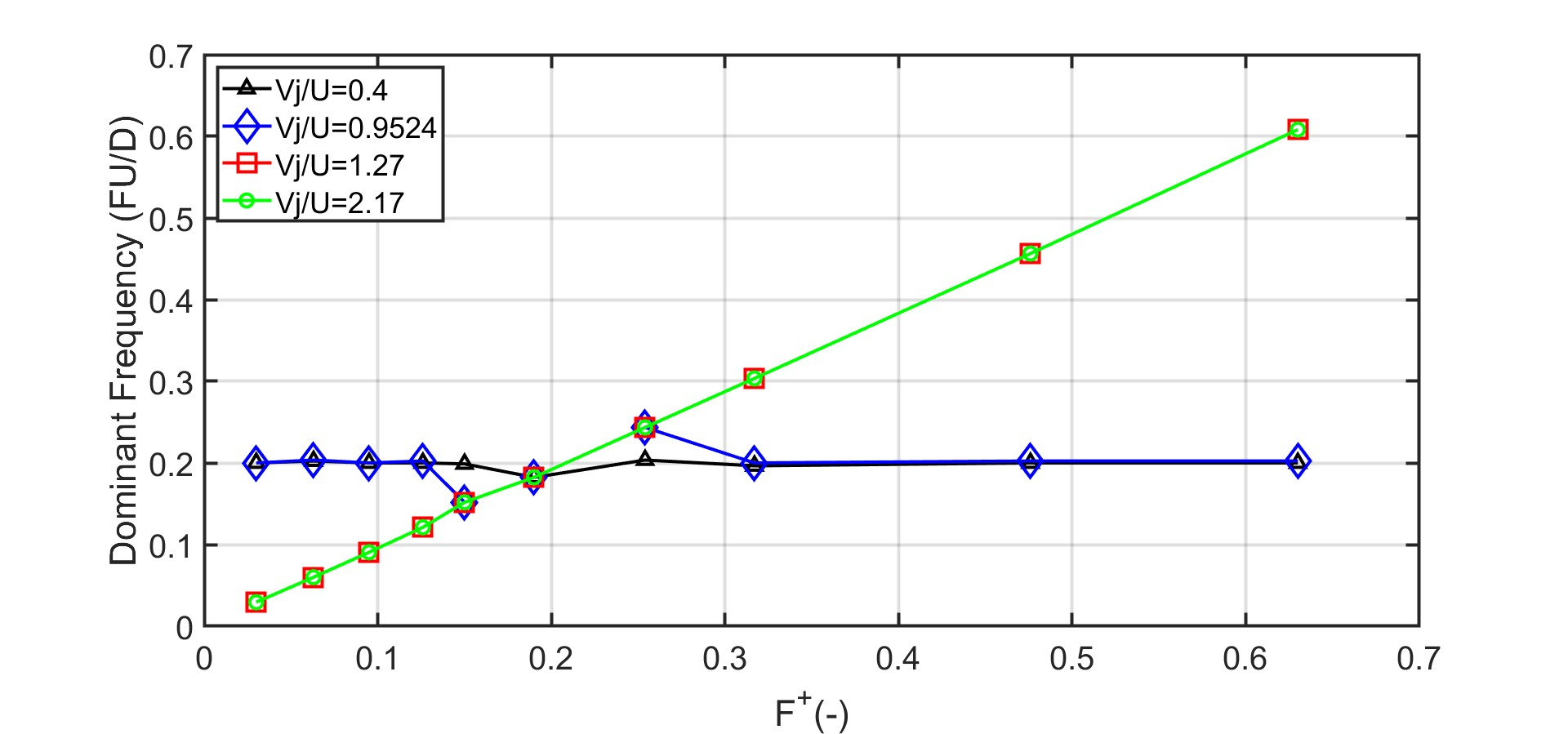}
    \caption{Dominant frequency observed in lift fluctuations depending on forcing frequency}
    \label{fig:image14}
\end{figure}

\section{Conclusions}
By implementing ASB \cite{garzozi2023wind}, alternating streams of pressurized air are ejected into the flow from the top and bottom of the cylinder, respectively. The rationale behind this technique is to excite the shear layer and periodically prevent flow separation, leading to amplified lift fluctuations. In this paper, we investigate the efficacy of ASB through experimentation in a low-speed wind tunnel over a wide range of parametric variations, exploring the Reynolds number of $Re = 8,000$ in detail. It was found that large velocity ratios and forcing frequencies close to the natural vortex shedding frequency give maximum lift fluctuations, which can enable the efficient conversion of the kinetic energy carried by the flow to electrical energy. Furthermore, lift fluctuations generated at a particular frequency of ASB are directly proportional to the momentum added to the flow. At threshold velocity ratios of around $V_\mathrm{R}=1.5$, ASB dominates the flow characteristics, meaning that the flow dynamics lock into the forcing frequency. The final two observations are also beneficial for traditional AFC to reduce aerodynamic drag and prevent any structural vibrations by preventing resonance frequency dominance in flows.

Based on the results presented in this paper, the introduction of AFC methods for the benefit of energy harvesting can be viewed as beneficial, especially with the ability to control and tune velocity ratios and pulsing frequencies to match environmental conditions, which contrasts with passive means of control. This ensures that energy is harnessed at the highest efficiency at the primary lock-in regime for VIVs, thus facilitating a bandwidth extension.

Future work needs to be dedicated to an oscillating cylinder setup to examine whether the mechanical motion of the bluff body affects the observed outcomes, potentially uncovering dynamic interactions between the cylinder's motion and the flow, as well as highlighting different modes of vortex shedding that may become prominent, hinting at the possibility of determining the frequency of pulsing required to be more involved.

\section*{Acknowledgements}
The authors gratefully acknowledge financial support from the Deutscher Akademischer Austauschdienst (German Academic Exchange Service) under project number 91897716.

%\cite{Yan2022}
%\cite{chaudhury2001vortex}
%\cite{islam2023flow}
%\cite{Blevins2001}
%\cite{WILLIAMSON2008713}
%\cite{lienhard1966synopsis}
%\cite{Strouhal1878}
%\cite{Landau1970}
%\cite{karman1912mechanismus}
%\cite{blevins1976fluid}
%\cite{feng1968measurement}
%\cite{Toebes1969}
%\cite{wilson1989vortex}
%\cite{wong1982comparative}
%\cite{sallet1970method}
%\cite{flatt1961history}
%\cite{Schlichting1968}
%\cite{seifert1998use}
%\cite{article}
%\cite{delaunay2001control}
%\cite{mathelin2002effect}
%\cite{glezer2002synthetic}
%\cite{bhattacharya2018optimum}
%\cite{lain2021understanding}
%\cite{10.1115/1.2957913}
%\cite{lu2022wind}
%\cite{li2009vertical}
%\cite{sun2020novel}
%\cite{zhang2017improving}
%\cite{xiong2016internal}
%\cite{garzozi2018pulsed}
%\cite{garzozi2023wind}
%\cite{greenblatt2012flow}
%\cite{guo2023effects}
%\cite{mei2021active}
%\cite{triyogi2009reducing}
%\newpage
\bibliographystyle{unsrt}  
\bibliography{references} 

\end{document}